\documentstyle[12pt]{article}
\def\uni{1\!\!1}
\def\bea{\begin{eqnarray}}
\def\eea{\end{eqnarray}}
\def\be{\begin{equation}}
\def\ee{\end{equation}}

\def\gam{\gamma}

\def\ldag{\lambda^\dagger}

\def\ss{\scriptstyle}

\newcommand\rf[1]{~(\ref{#1})}
\def\dx{\partial_x}
\def\dxa{\partial_{x_a}}
\def\dag{\dagger}
\def\nn{\nonumber}

\def\OR{|0_{\ss B},0_{\ss F}\rangle}

\def\la{\langle}

\def\ra{\rangle}

\def\m{\mu}
\def\n{\nu}
\def\g{\gamma}
\def\a{\alpha}
\def\b{\beta}
\def\ha{{\hat{\alpha}}}

\def\pr{\prime}
\def\ha{{\hat{\alpha}}}

\def\ldag{\lambda^\dagger}

\def\ss{\scriptstyle}

\def\min{|-\ra}
\def\pl{|+\ra}
\def\pmin{|\pm\ra}

\def\l{\lambda}
\newcommand{\NPB}[3]{Nucl.\ Phys. B#1 (#2) #3}

\newcommand{\PRD}[3]{Phys.\ Rev.\ D#1 (#2) #3}

\def\1ad{\mbox{\normalsize $^1$}}
\def\2ad{\mbox{\normalsize $^2$}}
\def\3ad{\mbox{\normalsize $^3$}}
\def\4ad{\mbox{\normalsize $^4$}}
\def\5ad{\mbox{\normalsize $^5$}}
\def\6ad{\mbox{\normalsize $^6$}}
\def\7ad{\mbox{\normalsize $^7$}}
\def\8ad{\mbox{\normalsize $^8$}}
\def\makefront{\vspace*{1cm}\begin{center}
\def\newtitleline{\\ \vskip 5pt}
{\Large\bf\titleline}\\
\vskip 1truecm
{\large\bf\authors}\\
\vskip 5truemm
\addresses
\end{center}
\vskip 1truecm
{\bf Abstract:}
\abstracttext
\vskip 1truecm}
\begin {document}
\rightline{NIKHEF 98-001}
\rightline{hep-th/yymmxxx}
\def\titleline{
Asymptotic Supergraviton States in
\newtitleline
Matrix Theory\footnote{Talk presented at the 
``31st International Symposium Ahrenshoop
on the Theory of Elementary Particles'' 
Buckow, September 2-6, 1997.}
}
\def\authors{
Jan Plefka and Andrew Waldron
}
\def\addresses{
NIKHEF\\ P.O. Box 41882, 1009 DB Amsterdam\\
The Netherlands
}
\def\abstracttext{
We study the Matrix theory from a purely canonical viewpoint.
In particular, we identify free particle asymptotic states of the model
corresponding to the 11D supergraviton multiplet along with the split of 
the matrix model Hamiltonian into a free and an interacting part.
Elementary quantum mechanical perturbation theory then yields an
effective
potential for these particles as an expansion in their inverse
separation.  We discuss how our scheme can
be used to compute the Matrix theory result for the 11D
supergraviton $S$ matrix and briefly  comment on non-eikonal and
longitudinal momentum exchange processes.
}
\makefront
\section{The model.}
Matrix theory~\cite{BFSS96} is the conjectured description of $M$ theory
in terms of a supersymmetric matrix model. At low energies and large
distances $M$ theory, by definition, reduces to 11D supergravity.
In this talk we explicitly construct asymptotic particle states in Matrix
theory to be identified with the 11D supergraviton multiplet and
study the scattering of these states.

The Hamiltonian of the Matrix theory
is that of ten dimensional $U(N)$
super Yang-Mills dimensionally reduced to $0+1$ dimensions~\cite{CH} and
arises from two disparate viewpoints.
On the one hand, it is the
regulating theory of the eleven dimensional supermembrane in light
cone gauge quantization \cite{dWHN88} and on the other,
it is the effective Hamiltonian describing the short distance properties
of $D0$ branes \cite{Polchinski95,Witten96,D0-branes}. Employing
the conjecture
of \cite{Susskind97}, the finite $N$ model is to be identified with
the compactification of a null direction of $M$ theory (henceforth
called the $-$ direction). The quantized total momentum of the
$U(N)$ system in this direction
is then given by $P_-=N/R$, where $R$ denotes the compactification
radius.

We shall be primarily interested in the $U(2)$ theory, studying
the Hilbert space of two supergravitons with momentum $P_-=1/R$ each.
The coordinates and Majorana spinors
of the transverse nine dimensional space then take values in the adjoint 
representation of $U(2)$, i.e.
\bea
X_\m &=& X^0_\m\,  i\,\uni + X^A_\m\, i\sigma^A \qquad \m=1,\ldots,9 \\
\theta_\alpha &=& \theta_\alpha^0\, i\, \uni + \theta_\alpha^A\,
i\sigma^A
\phantom{\sigma}
\qquad \alpha=1,\ldots, 16
\eea
where $\sigma^A$ are the Pauli matrices. We shall often employ a vector
notation for the $SU(2)$ part in which
$\vec{X}_\m= (X^1_\m,X^2_\m,X^3_\m)\equiv (X^A_\mu)$
and similarly for $\vec{\theta}$.

The Hamiltonian is then given by
\be
H=H_{\rm CoM}+
\frac{1}{2}\vec{P}_\m\cdot\vec{P}_\m
+\frac{1}{4}(\vec{X}_\m\times\vec{X}_\nu)^2+
\frac{i}{2}\vec{X}_\m\cdot
\vec{\theta}\g_\m\times\vec{\theta}
\label{ham}
\ee
where $H_{\rm CoM}=\frac{1}{2}RP^0_\m P^0_\m$ is the $U(1)$
centre of mass Hamiltonian. Note that we are using a real,
symmetric representation of the $SO(9)$ Dirac matrices in which the nine
dimensional charge conjugation matrix is equal to unity.

The Hamiltonian\rf{ham} is augmented by the Gauss law constraint
\be
\vec{L}=\vec{X}_\m\times\vec{P}_\m-
\frac{i}{2}\vec{\theta}\times\vec{\theta}\quad,\qquad
[L^A,L^B]=i\, \epsilon^{ABC} L^C\label{constraint}
\ee
whose action is required to vanish on physical states.

The task is now to identify the free asymptotic two-particle states of
the Hamiltonian\rf{ham} which describe the on-shell supergraviton
multiplet of eleven dimensional supergravity. This problem manifestly
factorises into a $U(1)$ centre of mass state and an $SU(2)$ invariant
state describing the relative dynamics of the particles.

\section{The centre of mass theory.}

The eigenstates of the free $U(1)$ centre of mass Hamiltonian  
$H_{\rm CoM}$
are
\be
|k_\mu;h_{\m\n},B_{\m\n\rho},h_{\m\ha}\ra_{_0}=
e^{ik_\m X^0_\m}
|h_{\m\n},B_{\m\n\rho},h_{\m\ha}\ra_{_0}\label{freesoln}
\ee
and possess transverse $SO(9)$ momentum $k_\m$ and on-shell
supergraviton polarisations\footnote{Note that the polarisation tensors 
$h_{\m\n}$, $B_{\m\n\rho}$ and $h_{\m\ha}$ correspond to {\it physical}
polarisations. The Matrix theory does away with unphysical
timelike and longitudinal polarisations at the price of manifest
eleven dimensional Lorentz invariance.} $h_{\m\n}$,
$B_{\m\n\rho}$ and $h_{\m\ha}$ (graviton, antisymmetric tensor and  
gravitino,
respectively).
The state $|h_{\m\n},B_{\m\n\rho},h_{\m\ha}\ra_{_0}$  is the
$\underline{44}\oplus\underline{84}\oplus\underline{128}$
representation of
the centre of mass spinor $\theta^0$ degrees of freedom.
The construction of this state is carried out in detail in~\cite{us} 
and allows the explicit calculation of the spin dependence of Matrix
theory supergraviton amplitudes.
In order to define the fermionic vacuum and creation and
annihilation operators
one performs a decomposition of the $SO(9)$ Lorentz algebra with
respect to an $SO(7)\otimes U(1)$ subgroup~\cite{dWHN88}.
This is done as follows.
Firstly split vector
indices $\m=(1,\ldots,9)$
as $(m=1,\ldots,7;8,9)$ so that an $SO(9)$ vector $V_\m$ may be
rewritten as $(V_m,V,V^*)$ where $V=V_8+iV_9$ and
$V^*=V_8-iV_9$.
For an $SO(9)$ spinor the same decomposition is made by complexifying,
in particular, for the canonical spinor variables we have
\bea
\l=\frac{\theta^0_++i\theta^0_-}{\sqrt{2}}\, , \,
\l^\dagger=\frac{\theta^0_+-i\theta^0_-}{\sqrt{2}}\label{napoleon},
\eea
where the subscript $\pm$ denotes projection by $(1\pm\g_9)/2$.
The canonical anticommutation relations are now
$
\{\l_\a,\l^\dagger_\b\}=\delta_{\a\b}$ where $\a,\b=1,\ldots,8$
and we define the fermionic vacuum $|-\ra$ by
$
\l|-\ra=0
$.
We denote the completely filled state by $\pl=\ldag_1\ldots\ldag_8\min$.
One finds then the following expansion for the supergraviton
polarisation state
\bea
|h_{\m\n},B_{\m\n\rho},h_{\m\ha}\ra_{_0}&=&h\min+\frac{1}{4}h_{m}\min_m
			+\frac{1}{16}h_{mn}\pmin_{mn}
			+\frac{1}{4}h^*_m\pl_m+h^*\pl\nn\\
			&&\hspace{-.8cm}-\frac{\sqrt{3}\, i}{8}
                 \left(\frac{}{}\!B_{mn}\min_{mn}
			+\frac{i}{6}B_m\pmin_m
                      +\frac{1}{6}B_{mnp}\pmin_{mnp}
          -B^*_{mn}\pl_{mn}\right)\nn\\
&&\frac{i}{\sqrt{2}}
\left(h_\a\min_\a-\frac{1}{2}h_{m\a}\min_{m\a}+\frac{1}{2}h^*_{m\a}\pl_{m\a}
  -h_\a^*\pl_\a\right)
\ .\nn\\&&\label{result}
\eea
The states in\rf{result} transform covariantly with respect to
$SO(7)\otimes U(1)$ and are defined in \cite{us}.

\section{Asymptotic states.}

Relative motions are described in the Matrix theory by the
constrained
$SU(2)$ quantum mechanical matrix theory defined above. However,
spacetime
is only an asymptotic concept in this theory. In particular diagonal
matrix configurations, i.e., those corresponding to Cartan generators
of $SU(N)$, span flat directions in the matrix model potential and
describe spacetime configurations~\cite{BFSS96}.
Transverse directions are described
by supersymmetric harmonic oscillator degrees of freedom, as we  
will see below.

Due to the gauge constraint\rf{constraint}
quantum mechanical wavefunctions must be invariant under $SU(2)$
rotations
so that there is no preferred Cartan direction.
To find asymptotic states corresponding to supergraviton
(i.e., spacetime) excitations in a gauge invariant way we proceed
as follows.
Let us suppose we wish to study states describing particles widely
separated in the (say) ninth spatial direction, then we may simply
declare the $SU(2)$ vector $\vec{X}_9$ to be large. The limit
$|\vec{X}_9|=\sqrt{\vec{X}_9\cdot\vec{X}_9}\rightarrow\infty$ is
$SU(2)$ rotation (and therefore gauge) invariant. We search for
asymptotic particle-like solutions in this limit.

To this end it is convenient to employ the (partial) gauge
choice~\cite{dWLN89} in
which one chooses a frame where $\vec{X}_9$ lies along the $z$-axis,
\be
X^1_9=0=X^2_9\, . \label{ed}
\ee
Calling $X_9=(0,0,x)$ and $\vec{X}_a=(Y^1_a,Y^2_a,x_a)$ (with
$a=1,...,8$) the Hamiltonian in this frame
then is $H=H_{\rm V}+H_{\rm B}+H_{\rm F}+H_4$ where\footnote{The spinors 
$\tilde{\theta}$ are built from $\theta^1$ and $\theta^2$ by
complexification and
a ${\rm spin}(9)$ rotation (see \cite{us}).
Note that $r^2\equiv x_a x_a+x^2$.}
\bea
H_{\rm V}&=& -\frac{1}{2x}\, (\dx)^2x -\frac{1}{2}\, (\dxa)^2 \\
H_{\rm B}&=& -\frac{1}{2}\, (\frac{\partial}{\partial {Y_a^I}})^2
+ \frac{1}{2}\, r^2\, Y_a^I\, Y_a^I \label{tom}\\
H_{\rm F}&=& r\, \tilde{\theta}^\dag\, \gam_9\, \tilde{\theta}\,\\
H_4&=& \mbox{``rest''} .
\eea
The sum of the  Hamiltonians $H_{\rm B}$ and
$H_{\rm F}$ is that of a supersymmetric
harmonic oscillator with frequency $r$ and describes excitations
transverse to the flat directions. Particle motions in the flat
directions
correspond to the Hamiltonian $H_{\rm V}$ whereby we interpret
the Cartan variables $x_\m=(x_a,x)$ asymptotically
as the $SO(9)$ space coordinates.

The Hilbert space may be treated as a ``product'' of superoscillator
degrees of freedom and Cartan wavefunctions depending on $x_\m$ and the
third component of $\vec{\theta}$ via the identity
\bea
H=\sum_{m,n}|m\ra\, \la m|H|n\ra\, \la n|
\eea
where $\{|n\ra\}$ denote the complete set of eigenstates of $H_{\rm  
B}$ and
$H_{\rm F}$. Since the frequency $r$ of the superoscillators is
coordinate
dependent, operators $\partial/\partial x_\m$ do not commute with $|n\ra$
so that this ``product'' is not direct.
This construction allows us to study an ``effective''
Hamiltonian
$H_{mn}(x_\mu,\partial_{x_\mu}, \theta^3)=\la m|H|n\ra$
for the Cartan degrees of freedom pertaining to
asymptotic spacetime.
In particular the free Hamiltonian is given by
the diagonal terms\footnote{We subtract terms $c_n/r^2$ to ensure
the correct
asymptotic behaviour of the interaction Hamiltonian.
A detailed explanation of this point may be found in \cite{us}.}
\be
H_0=\sum_n \, |n\rangle\, \langle n|\left( H_{\rm V}+H_{\rm B}+H_{\rm F} 
\, -\,\frac{c_n}{r^2}\right) |n\rangle\,
\langle n|\label{free}
\ee
and the interaction Hamiltonian then reads $H_{\rm Int}=H-H_{\rm
CoM}-H_0$.
Since supersymmetric harmonic oscillator zero point energies vanish,
eigenstates of\rf{free} are
\be
|k_\m;h_{\m\n},B_{\m\n\rho},h_{\m\ha}\ra=\frac{1}{x}e^{ik_\m x_\m}
|h_{\m\n},B_{\m\n\rho},h_{\m\ha}\ra\otimes|0_B,0_F\ra\label{solutions}
\ee
where $|0_B,0_F\ra$ is the supersymmetric harmonic oscillator vacuum.
These states satisfy the correct free particle dispersion relation
\be
H_0|k_\m;h_{\m\n},B_{\m\n\rho},h_{\m\ha}\ra
=\frac{1}{2}k_\m k_\m\, |k_\m;h_{\m\n},B_{\m\n\rho},h_{\m\ha}\ra\ .
\ee
Here, the supergraviton polarisation multiplet
$|h_{\m\n},B_{\m\n\rho},h_{\m\ha}\ra$ is built from the
$\underline{44}\oplus\underline{84}\oplus\underline{128}$
representation of $\theta^3$ as in\rf{result}.

Therefore, upon taking the direct product of an asymptotic
state\rf{solutions}
with a centre of mass eigenstate\rf{freesoln}
\be
{|1,2\ra}=
{e^{i k_\mu^{\rm{tot}}X_\mu^0}}\, {\frac{1}{x}
e^{i k_\mu^{\rm {rel}}x_\mu}}\otimes \OR\otimes
|h^1_{\m\n},B^1_{\m\n\rho},h^1_{\m\ha}\ra_{\theta^0+\theta^3}
 \otimes |h^2_{\m\n},B^2_{\m\n\rho},h^2_{\m\ha}
\ra_{\theta^0-\theta^3}
\ee
one obtains a state
describing a pair of supergravitons widely separated in the ninth
spatial direction. Its interactions, which die off as
$x\rightarrow\infty$,  are governed by $H_{\rm Int}$.

\section{Scattering amplitudes.}

The $2\longrightarrow 2$ supergraviton scattering amplitude is then
obtained by elementary quantum mechanical scattering theory as
\bea
\lim_{T\rightarrow\infty}\, \la1^\prime,2^\prime| e^{-i H\, T}|1,2\ra&=&
\delta(k_\mu^{\prime \rm{tot}}-k_\mu^{\rm{tot}}) \, \int 4\pi x^2\,
d^9 x_\mu \\ \nonumber&&
\frac{\textstyle e^{-ik_\m^{\prime\,{\rm rel}} x_\m}}{\textstyle x}\,
\la {\cal H}^{1^\pr},{\cal H}^{2^\pr}|\, {H_{\rm Eff}}
(x_\m,\partial_\m,
\theta^3_\ha)\,
|{\cal H}^1,{\cal H}^2\ra
\, \frac{\textstyle e^{ik_\m^{\rm rel} x_\m}}{\textstyle
x}\label{amplitude}
\eea
where we have introduced $|{\cal H}^1,{\cal H}^2\ra=
|h^1_{\m\n},B^1_{\m\n\rho},h^1_{\m\ha}\ra_{\theta^0+\theta^3}
 \otimes |h^2_{\m\n},B^2_{\m\n\rho},h^2_{\m\ha}
\ra_{\theta^0-\theta^3}$ and similarly for
$|{\cal H}^{1^\pr},{\cal H}^{2^\pr}\ra$. The leading (Born) approximation
to the ``effective'' Cartan Hamiltonian ${H_{\rm Eff}}$ is given by
\be
{H_{\rm Eff}}^{(1)}(x_\mu,\partial_{x_\mu}, \theta^3_\ha)=
\la 0_B,0_F|\, H_{\rm Int} | 0_B,0_F\ra
\ee
and higher order contribution are obtained from the Lippman-Schwinger
expansion
\bea
{H_{\rm Eff}}(x_\m,\partial_\m,\theta^3_\ha)
&=&\la 0_B,0_F| {H_{\rm Int}}|0_B,0_F\ra + \la 0_B,0_F|{H_{\rm Int}}
{\textstyle\frac{1}{E-{H_0}+i\epsilon}} {H_{\rm Int}}|0_B,0_F\ra
\nonumber\\
&&+ \la 0_B,0_F|{H_{\rm Int}}
{\textstyle \frac{1}{E-{H_0}+i\epsilon}} {H_{\rm Int}}
{\textstyle\frac{1}{E-{H_0}+i\epsilon}}
{H_{\rm Int}}|0_B,0_F\ra+\ldots\label{Lipp}
\eea
which due to the scaling behaviours $H_{\rm Int}\sim{\cal O}(x^{-1/2})$
and $H_0\sim{\cal O}(x)$ turns out to be an expansion in $1/x$ the  
inverse
separation of the two supergravitons \cite{us}.

The leading term of $H_{\rm Eff}$ is on dimensional grounds of order 
$1/r^2$ and receives contribution at first and second order perturbation
theory in the sense of\rf{Lipp}. An explicit computation
\cite{us} shows
that these two contributions precisely cancel\footnote{Similar  
computations
have been performed in a different context in\cite{SS}.}.
This supersymmetric cancellation is in accordance with the two loop
semiclassical background field path integral calculation of
\cite{BBPT97}
and yields a strong test of our proposal. Higher order contributions
should then capture the revered $v^4/r^7$ potential for $D0$
particles~\cite{DKPS97}.

Let us stress at this point, however, that the amplitudes\rf{amplitude}
are restricted to the eikonal kinematical regime (i.e. high energy,
straight line), as the
asymptotic in- and outgoing supergraviton pairs are widely separated 
in the same (in this case  9th) spatial direction. Scattering amplitudes
at arbitrary angles ($\sigma_{\mu\nu}$) may be obtained by performing an
$SO(9)$ rotation of the outgoing state, i.e.
\be
{\la 1^\pr, 2^\pr|}\exp(i\, \sigma_{\mu\nu}\,
{L^{\mu\nu}_{\rm SO(9)}})\,
\exp(-i H\, T) {|1,2\ra}
\ee
where ${L^{\mu\nu}_{\rm SO(9)}}$ denotes the generator of $SO(9)$.

The $2 \longrightarrow 1$ supergraviton scattering channel of the
$U(2)$ Matrix theory hinges on the knowledge of the zero-energy  
groundstate
$|\mbox{GS}\ra$ of the $SU(2)$ supersymmetric quantum mechanics
(which exists, according to~\cite{SS,PR97}).
The ``1'' supergraviton state with $P_-=2/R$ is then given by the direct
product of the $U(1)$ centre of mass state $|k^{1^\pr}_\mu,
{\cal H}^{1^\prime}\ra_0$
with $|\mbox{GS}\ra$. Therefore the $2 \longrightarrow 1$ amplitude reads
\bea
\lim_{T\rightarrow\infty}\, \la1^\prime| e^{-i H\, T}|1,2\ra&=&
{}_{{}_0}\la k_\mu^{1^\pr},{\cal H}^{1^\pr}|\otimes {\la
{\rm GS}|}\,
\exp(-iHT)\,
|k_\mu^{\rm 1},k_\mu^{\rm 2};{\cal H}^{\rm 1},
{\cal H}^{\rm 2}\,\rangle\nn\\
&=&{}_{{}_0}\la k_\mu^{1^\pr},{\cal H}^{1^\pr}|\otimes {\la
{\rm GS}|}\,
k_\mu^{\rm 1},k_\mu^{\rm 2};{\cal H}^{\rm 1},
{\cal H}^{\rm 2}\,\rangle \,
\eea
since $H|{\rm GS}\ra=0$. Exact knowledge of the state $|{\rm GS}\ra$ would
yield us the complete non-perturbative answer for this process
involving longitudinal momentum exchange. Recently there has been
some progress towards uncovering the structure of the ground state
\cite{H97,HS97}.

\bigskip
\leftline{\bf Acknowledgements}

\medskip
\noindent J. Plefka thanks the organizers for a stimulating symposium.


\end{document}